\documentclass[12pt]{spieman}  
\usepackage{amsmath,amsfonts,amssymb}
\usepackage{graphicx}
\usepackage{setspace}
\usepackage[figure,figure*]{hypcap}
\usepackage{chemformula}
\usepackage{subfigure}

\def\kep{\emph{Kepler}~}

\def\mps{m\,s$^{-1}$~}

\def\cmps{cm\,s$^{-1}$~}
\def\cmpsp{cm\,s$^{-1}$}
\def\ln{L\ch{N_2}~}
\def\lnp{L\ch{N_2}}
\newcommand{\unit}[1]{\ensuremath{\, \mathrm{#1}}}

\title{Ultra-Stable Environment Control for the NEID Spectrometer: Design and Performance Demonstration}

\author[a,*]{Paul Robertson}

\author[b]{Tyler Anderson}

\author[c,b,d]{Gudmundur Stefansson}

\author[b]{Frederick R. Hearty}

\author[b]{Andrew Monson}

\author[b,d]{Suvrath Mahadevan}

\author[b]{Scott Blakeslee}

\author[e]{Chad Bender}

\author[b,d]{Joe P. Ninan}

\author[b]{David Conran}

\author[b]{Eric Levi}

\author[b]{Emily Lubar}

\author[f]{Amanda Cole}

\author[g]{Adam Dykhouse}

\author[b,d]{Shubham Kanodia}

\author[b]{Colin Nitroy}

\author[h]{Joseph Smolsky}

\author[i]{Demetrius Tuggle}

\author[j]{Basil Blank}

\author[k]{Matthew Nelson}

\author[l]{Cullen Blake}

\author[m,n]{Samuel Halverson}

\author[o]{Chuck Henderson}

\author[e]{Kyle F. Kaplan}

\author[p]{Dan Li}

\author[q,r]{Sarah E. Logsdon}

\author[s]{Michael W. McElwain}

\author[p]{Jayadev Rajagopal}

\author[b,d]{Lawrence W. Ramsey}

\author[t]{Arpita Roy}

\author[u]{Christian Schwab}

\author[v]{Ryan Terrien}

\author[b,d]{Jason T. Wright}

\affil[a]{The University of California, Irvine, Department of Physics \& Astronomy, 4129 Frederick Reines Hall, Irvine, CA, USA, 92697}

\affil[b]{The Pennsylvania State University, Department of Astronomy \& Astrophysics, University Park, PA, USA}

\affil[c]{NASA Earth and Space Science Fellow}

\affil[d]{The Pennsylvania State University, Center for Exoplanets \& Habitable Worlds, University Park, PA, USA}

\affil[e]{The University of Arizona, Department of Astronomy and Steward Observatory, Tucson, AZ, USA}

\affil[f]{The Pennsylvania State University, College of Engineering, University Park, PA, USA}

\affil[g]{The Pennsylvania State University, Department of Physics, University Park, PA, USA}

\affil[h]{The Massachusetts Institute of Technology, Laboratory for Nuclear Science, Cambridge, MA, USA}

\affil[i]{The Ohio State University, Department of Astronomy, Columbus, OH, USA}

\affil[j]{PulseRay Inc., Beaver Dams, NY, USA}

\affil[k]{The University of Virginia, Department of Astronomy, Charlottesville, VA, USA}

\affil[l]{The University of Pennsylvania, Department of Physics \& Astronomy, Philadelphia, PA, USA}

\affil[m]{NASA Sagan Fellow}

\affil[n]{The Massachusetts Institute of Technology, Department of Physics, Cambridge, MA, USA}

\affil[o]{Cornell University, Ithaca, NY, USA}

\affil[p]{National Optical Astronomy Observatory}

\affil[q]{NASA Postdoctoral Program Fellow}

\affil[r]{NASA Goddard Space Flight Center}

\affil[s]{Exoplanets and Stellar Astrophysics Laboratory, NASA Goddard Space Flight Center, Greenbelt, MD USA}

\affil[t]{The California Institute of Technology, Department of Astronomy, Pasadena, CA, USA}

\affil[u]{Macquarie University, Department of Physics \& Astronomy, Sydney, Australia}

\affil[v]{Carleton College, Department of Physics \& Astronomy, Northfield, MN, USA}

\begin{document} 
\maketitle

\begin{abstract}
Two key areas of emphasis in contemporary experimental exoplanet science are the detailed characterization of transiting terrestrial planets, and the search for Earth analog planets to be targeted by future imaging missions.  Both of these pursuits are dependent on an order-of-magnitude improvement in the measurement of stellar radial velocities (RV), setting a requirement on single-measurement instrumental uncertainty of order 10 cm/s.  Achieving such extraordinary precision on a high-resolution spectrometer requires thermo-mechanically stabilizing the instrument to unprecedented levels.  Here, we describe the Environment Control System (ECS) of the NEID Spectrometer, which will be commissioned on the 3.5\,m WIYN Telescope at Kitt Peak National Observatory in 2019, and has a performance specification of on-sky RV precision $< 50$ cm/s.  Because NEID's optical table and mounts are made from aluminum, which has a high coefficient of thermal expansion, sub-milliKelvin temperature control is especially critical.  NEID inherits its ECS from that of the Habitable-zone Planet Finder (HPF), but with modifications for improved performance and operation near room temperature.  Our full-system stability test shows the NEID system exceeds the already impressive performance of HPF, maintaining vacuum pressures below $10^{-6}$ Torr and an RMS temperature stability better than $0.4$ mK over 30 days.  Our ECS design is fully open-source; the design of our temperature-controlled vacuum chamber has already been made public, and here we release the electrical schematics for our custom Temperature Monitoring and Control (TMC) system.   
\end{abstract}


{\noindent \footnotesize\textbf{*}Paul Robertson,  \linkable{paul.robertson@uci.edu} }


\section{Introduction}
\label{sec:intro}  
The field of exoplanet discovery and characterization has advanced to the point of observing exoplanets analogous to the rocky planets of the inner Solar system.  The \kep mission \cite{borucki10} and its extended campaign K2 \cite{howell14} have discovered thousands of exoplanets, most of which have radii smaller than that of Neptune.  The TESS spacecraft \cite{ricker15} has recently begun an all-sky survey for the nearest transiting exoplanets, which will be optimal targets for follow-up characterization with facilities such as the James Webb Space Telescope and 30\,m-class ground-based telescopes.

The nearby transiting exoplanets discovered by TESS offer an unprecedented opportunity to constrain their interior structures, composition, and atmospheric dynamics.  However, the scientific yield of this endeavor is highly dependent on the availability of ultra-precise Doppler spectroscopy.  Comparative planetology of Earth- and super-Earth-mass exoplanets is dependent on determining masses \emph{and} radii of the sample population to better than 20 percent, which requires radial velocity (RV) precision better than the 1 \mps standard established by instruments such as HARPS \cite{pepe03} and HIRES \cite{vogt94}.

Precision radial velocimetry is achieved through one of two spectrometer design concepts.  The first, described in detail by Refs.~\citenum{valenti95} and \citenum{butler96}, places a temperature-controlled cell of gas in the optical path of the instrument.  The gas is selected so as to superimpose a series of thousands of narrow, optically thin absorption lines over the observed stellar spectrum.  The lines have precisely known, stable wavelengths, and may therefore serve as a wavelength fiducial.  Furthermore, because the absorption lines share a common optical path as light from the target, any changes to the instrument profile (IP) caused by instability in the spectrometer will be reflected in the observed absorption line profiles, and controlled for.  The derived radial velocity is then computed as a free parameter in the forward model of the star-plus-absorption cell spectrum.  This technique is often referred to as the ``iodine method,'' due to the prevalence of molecular iodine (I$_2$) as a reference gas in optical spectrometers.  The iodine technique is broadly applicable in the sense that most general-purpose high resolution spectrometers may be easily retrofitted with a gas cell without additional modification or stabilization.  On the other hand, the iodine method is limited primarily by the fact that Doppler information may only be extracted from the waveband covered by the I$_2$ absorption spectrum, and additionally by algorithmic complexity in forward-modeling the star-I$_2$ spectrum.

Most new instruments dedicated to precision Doppler spectroscopy typically adopt the second method of achieving high RV precision.  Sometimes referred to as the ``cross-correlation method'', this technique requires stabilizing the full opto-mechanical train of a fiber-fed spectrometer to an extreme degree, thereby permanently fixing the IP \cite{baranne96}.  Absent the need to track IP variations, a photonic source such as a hollow cathode lamp or a laser frequency comb (LFC) may then be observed simultaneously with the starlight on a second fiber, providing wavelength calibration and measuring instrument drift across the spectrometer's entire bandpass.  Because the ensuing Doppler extraction model involves significantly fewer free parameters than the absorption cell technique, precise RVs may be determined at lower S/N values, reducing the telescope apertures and exposure times required to collect RV data.  Stabilized spectrometers can also extract RVs from outside the $\sim500-600$ nm I$_2$ absorption band, where there exists significant Doppler information content \cite{bouchy01} for both Sunlike stars (blueward of 500 nm) and the cooler M dwarfs (redward of 600 nm).  The cross-correlation technique therefore provides access to much more Doppler information content, and simplifies computational analysis, at the expense of placing much more stringent constraints on the physical stability of the instrument.  While simultaneous calibration can be used to monitor and correct drifts associated with slowly-evolving error sources such as the glass-crystal phase change of the \'echelle grating's Zerodur substrate \cite{bayer87}, the algorithms used to calculate both instrument drifts and on-sky RVs must assume the IP is stable.  If this assumption is violated due to environmental instability, uncorrectable errors may be introduced.

We have developed two stabilized, fiber-fed, high resolution spectrometers to facilitate detection and characterization of low-mass exoplanets around nearby stars.  The first is the Habitable-zone Planet Finder (HPF)\cite{mahadevan14}, a near-infrared (NIR) spectrometer  on the 10\,m Hobby-Eberly Telescope (HET) at McDonald Observatory.  The instrument's NIR wavelength coverage, combined with the 10\,m HET aperture, will enable it to detect Earth-mass planets in the liquid-water habitable zones (HZs)\cite{kopparapu13} of nearby fully-convective M dwarfs.  

Much of the design heritage of HPF was adapted to develop the optical-bandpass NEID spectrometer\cite{schwab16} for the 3.5\,m WIYN Telescope at Kitt Peak National Observatory.  NEID will combine the extreme environment stabilization of HPF with a high-performance, low-risk optical design, and a laser frequency comb wavelength calibrator.  NEID will be available for shared risk science beginning in Fall 2019, providing ultra-precise ($\sigma_{\textrm{RV}} < 50$ \cmpsp) RV capability on a publicly-accessible instrument in time to constrain masses of exoplanets discovered by TESS.

\begin{figure*}
\begin{center}
\includegraphics[width=\textwidth]{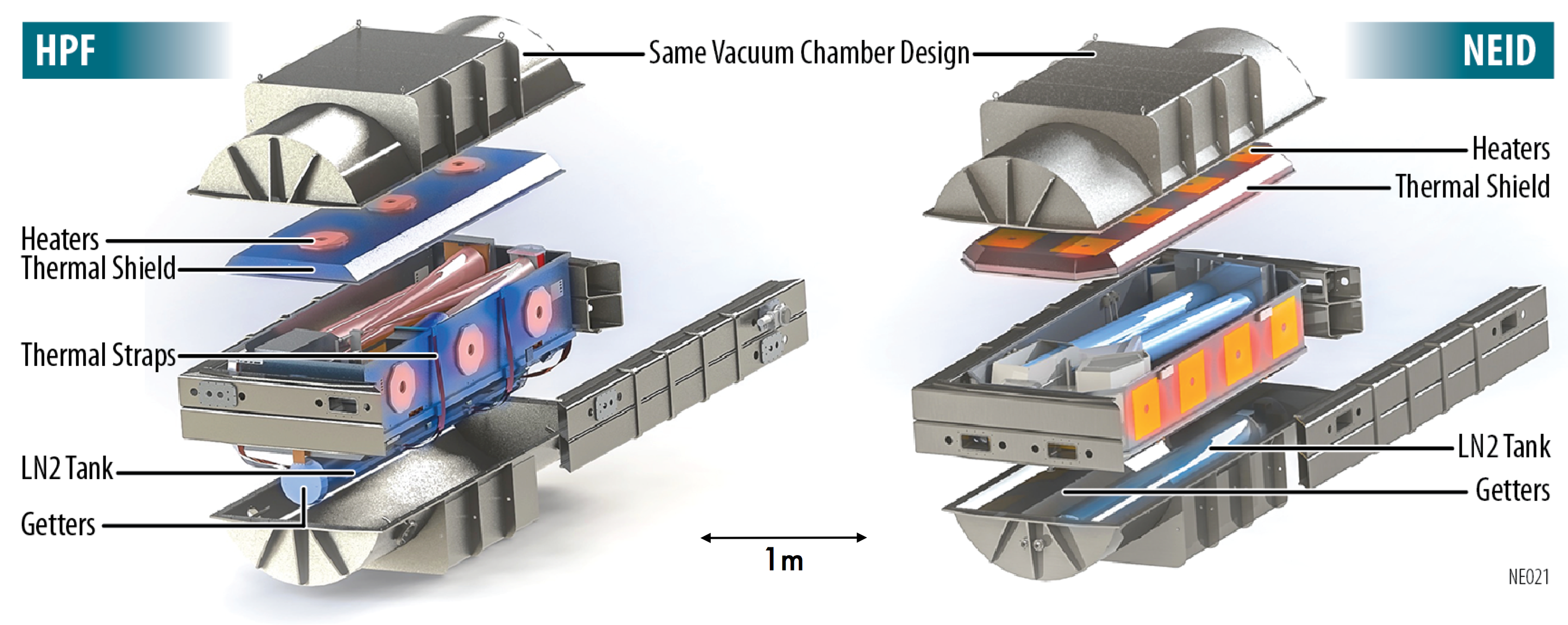}
\caption{\footnotesize Design schematics for the HPF (\emph{left}) and NEID (\emph{right}) vacuum chambers and radiation shields.}
\label{fig:HPFvNEID}
\end{center}
\end{figure*}

In order to achieve $\sigma_{\textrm{RV}} < 50$ \cmps upon delivery, and provide a reasonable path towards 10 \cmps precision, we have assembled a detailed error budget for NEID \cite{halverson16}.  This error budget requires, at minimum, that the spectrometer optics be maintained at $<1~\mu$Torr absolute pressure and $<1$ milliKelvin relative temperature stability over an extended period.  Our environmental stability requirements are stringent, even relative to other Doppler spectrometers, because our optical bench and optics mounts are all made of 6061-T6 aluminum.  Our choice of material offers ready availability, ease of machining, and cost effectiveness in exchange for a high coefficient of thermal expansion (CTE) relative to more exotic materials such as Zerodur (a la the Keck Planet Finder)\cite{gibson16}, Cordierite (a la the Infrared Doppler Instrument for Subaru)\cite{kotani2014}, or Invar (iLocater)\cite{crass16}.  The high conductivity of aluminum is used to minimize the ability of the optical bench and optics mounts to maintain thermal gradients.  The aluminum optical mounts are coupled to the optical bench through precision-machined interface pads that are securely bolted to ensure good thermal transfer.  The overall thermal design of the blackbody cavity in which all optical components are enclosed yields almost uniform temperatures of components at steady state, and quickly dissipates any gradients produced during short-term upset conditions.  Thus, long term stable and uniform temperatures are a necessity for instruments blazing a path to 10 \cmps RV precision.  The increased CTE of the aluminum bench and mounts translates directly to a greater temperature sensitivity in our Doppler error budget, since the optics will experience relative position shifts due to their mounting expanding or contracting with changes in temperature.

The specific 1 mK thermal stability requirement for NEID is driven by tolerances in the dispersive optics, their mounts, and the optical bench.  Principal among the various thermal factors is the change of groove spacing on the \'echelle grating with temperature.  Our grating is made from Zerodur Expansion Class 0, which yields 1.2 \cmps RV error per 1 mK change; Class 1 and 2 would produce $2.5 \times$ and $5 \times$ larger errors, respectively, so they were chosen against.  Additional thermal errors of this same order (1 \cmps per mK) are associated with temperature changes in the prism cross disperser and optical bench.  Combined in quadrature, these three sources of error---assuming 1 mK is the current best estimate of performance---yield a thermo-mechanical error consistent with other sources such as Zerodur glass-crystal phase change, thermal transients in the CCD detector during integration, and detector charge transfer efficiency.  As long as these errors remain consistent and reproducible, they can be eliminated in large part by calibration.

In this paper, we present the design and performance demonstration of NEID's environment control system (ECS).  The NEID ECS is conceptually similar to that of HPF, which has already demonstrated sub-milliKelvin temperature stability in both warm ($T\sim300$ K) and cold ($T\sim180$ K) operation (see \citenum{stefansson16}, hereafter S16).  The NEID design includes a number of improvements and optimizations from the HPF baseline, some of which were propagated to the final HPF ECS installation as well.  In particular, the custom electronics used to monitor and control the temperature of the spectrometer's active volume were completely overhauled, yielding greatly improved performance.  Here, we will briefly review the overall HPF/NEID ECS design concept, while focusing on the modifications implemented since the demonstrations described in S16 and the full-scale NEID performance test.

The paper is organized as follows.  In Section \ref{sec:concept}, we review the top-level design concept for our environment control system.  In Section \ref{sec:differences}, we highlight the specific differences between the HPF and NEID systems.  Section \ref{sec:electronics} details the design of our custom electronics for monitoring and controlling the temperature inside the NEID vacuum chamber.  We describe the setup of our 30-day environmental stability test in Section \ref{sec:test}, and present the results of that experiment in Section \ref{sec:data}.  Finally, we discuss and contextualize our results in Section \ref{sec:discussion}.

\section{ECS Design Concept}
\label{sec:concept}

The goal of the HPF/NEID environment control system is to achieve maximal pressure and temperature stability while minimizing thermo-mechanical transients and vibrations that might perturb ultra-precise RV measurements.  Because any servicing or modification of a Doppler spectrometer incurs the risk of introducing zero-point offsets to RV time series (as observed for the Hamilton Spectrograph\cite{fischer14}), the ECS must be able to maintain stable conditions over years with minimal replenishment or maintenance.  The design of our ECS is described in great detail in S16, and therefore will not be repeated here.  Rather, we briefly highlight the core elements of our ECS concept, and refer the interested reader to S16 for a full account, including open-source schematics for our vacuum chamber.

The ECS eliminates RV errors caused by index-of-refraction variability and facilitates excellent temperature stability by placing the entire optical train in a high-quality vacuum.  Our vacuum chamber is first evacuated to $P\sim10^{-4}$ Torr using conventional dry-scroll pumps.  Then, we inject liquid nitrogen (\lnp) into a tank mounted within the chamber.  The \ln cools  activated charcoal getters, which at cryogenic temperatures absorbs any remaining air in the system, resulting in stable pressures near $P < 10^{-6}$ Torr.  In addition to cryo-pumping the vacuum chamber, the \ln tank provides cooling to NEID's CCD detector via a cold finger extending from the tank through the optical bench.  Thus, our ECS provides high-quality vacuum and cooling without continuously running vacuum pumps or electrical cryocoolers, each of which would otherwise create unacceptable levels of vibration for our stability requirements.  

The upper and lower lids of the vacuum chamber are sealed with Viton O-rings, resulting in a total O-ring length of 20 m.  This corresponds to a total gas load of $2 \times 10^{-5}$ Torr/L/s due to O-ring permeation\cite{stefansson16}, assuming a standard Viton permeation rate of $2.5 \times 10^{-8}$ Torr/L/s/inch\cite{ohanlon03}.  Based on this leak rate, as well as experience from similar vacuum chambers used by APOGEE\cite{blank10} and HPF\cite{stefansson16}, our charcoal getters are sized to maintain the vacuum pressures nominally at $P < 10^{-6}$ Torr for 3 years without saturating, so internal instrument maintenance requirements should be minimal.

In Figure \ref{fig:HPFvNEID}, we show the exploded schematics of the vacuum chambers for both HPF and NEID, so that readers may compare and contrast the differences between the two instruments, which are discussed in detail in Section \ref{sec:differences}.

As noted in S16, the stainless steel alloy used in the construction of our vacuum chamber outgasses hydrogen at the operating vacuum pressures of our system. Activated charcoal does not absorb hydrogen, so our ECS also includes a NEXTorr D-100 non-evaporative hydrogen getter and sputter ion pump from SAES Getters. This device eliminates the slow vacuum degradation we would otherwise observe due to hydrogen outgassing.  Before deployment at WIYN, we will upgrade to a D-200 getter for greater capacity and thus longer hold times.

Within the vacuum chamber, the spectrometer's optical bench and its components are suspended on stainless steel rods mounted via ball joints to steel blocks welded to the inside of the chamber.  Contact points between the rods and the bench/blocks are buffered with G10 spacers, which are thermally non-conductive.  The instrument's temperature is therefore regulated almost entirely radiatively, since heat transfer pathways through convection and conduction are eliminated by the vacuum and non-conductive bench suspension, respectively.  

To maintain constant temperature, the NEID optical bench is encased within an actively-controlled aluminum radiation shield, which is held to the optical bench with non-conductive G10 blocks. The surface of the NEID radiation shield is divided into 28 zones that are each independently actively controlled using our custom Temperature Monitoring and Control (TMC) system.  Each zone is controlled by a $12" \times 12"$ adhesive Kapton resistive heater produced by Birk Manufacturing, run on a closed-loop proportional-integral-derivative (PID) feedback cycle. 

\begin{figure}
\centering
\includegraphics[width=0.4\textwidth]{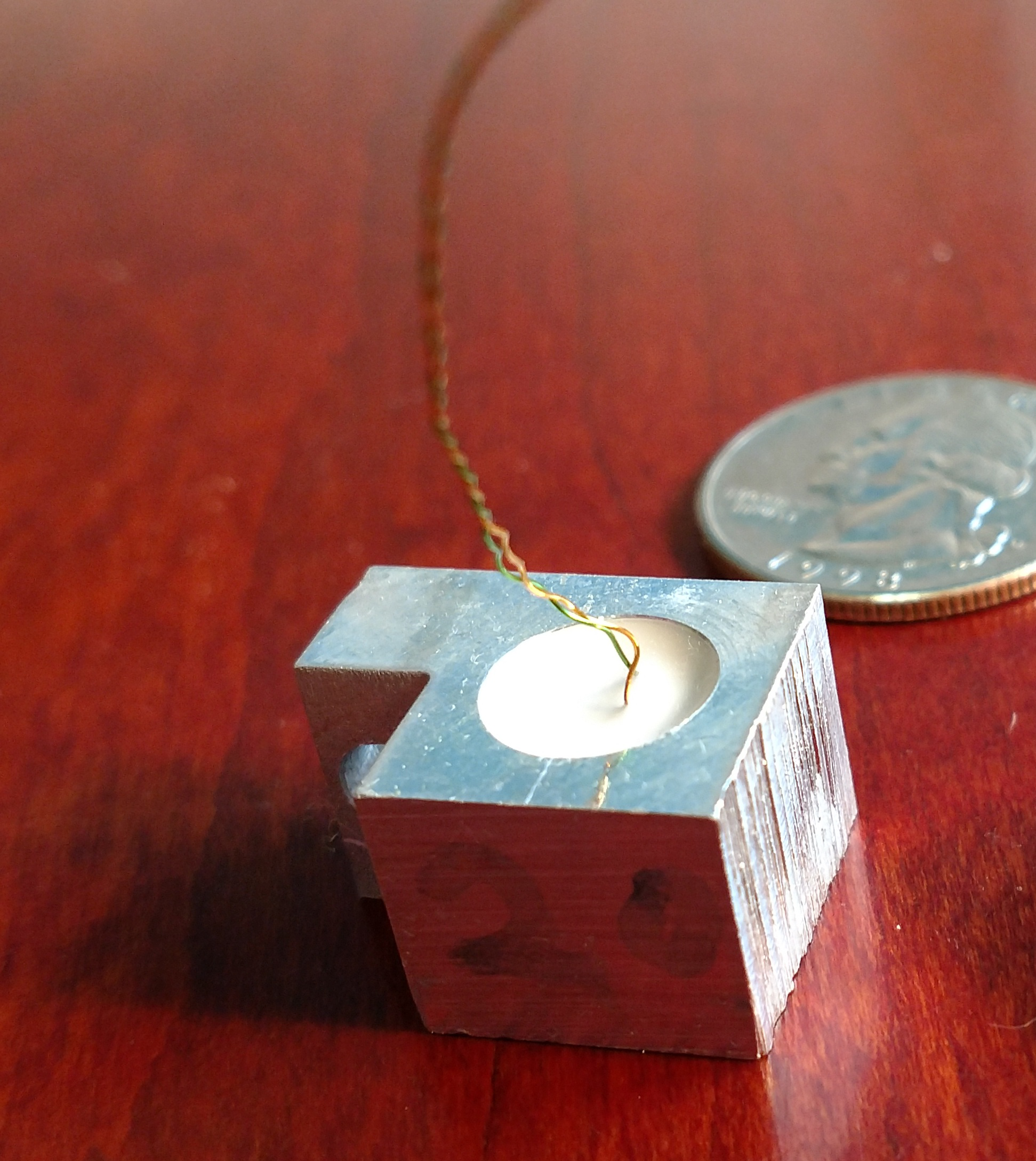}
\caption{\footnotesize A control thermometer for the NEID TMC.  A 2N2222 transistor, operated as a diode, is potted in a small aluminum block with thermal epoxy.}
\label{fig:2N2222}
\end{figure}

Input telemetry to each PID loop is provided by a custom thermometer (hereafter ``control thermometer'') based on a 2N2222 transistor operated as a diode.  The control thermometers are potted in a small block of aluminum with thermal epoxy to facilitate ease of mounting while maintaining thermal conductivity.  We show a potted control thermometer in Figure~\ref{fig:2N2222}.  When used as a thermometer in this way, the 2N2222 devices compare favorably to commercial-off-the-shelf temperature probes.  They are inexpensive ($\sim 7$ USD each), reliable\cite{nasa1978}, and exhibit $\sim 0.1$ mK measurement stability.  While this performance is exceeded by some commercial devices, our decision to use custom thermometers and control electronics is informed both by measurement precision and the costs associated with controlling so many separate channels.  For commercial systems that meet our performance requirements, the cost of a single control channel---including a thermometer, monitoring electronics, and power output---ranges from 1500-2000 USD or more, and typically offers lower precision than demonstrated here.  The hardware cost of our system is of order 250 USD per control channel, and requires far less space in the instrument's electronics cabinet.

Each control thermometer is paired with a separate, high-precision PT-103 resistive thermometer (hereafter ``monitoring thermometer'') from LakeShore Cryogenics and monitored by a microK 250 thermometry bridge from Isotech. The monitoring thermometers provide absolute calibration for the control thermometers, as well as offering an independent fiducial to ensure our temperature control system is performing as expected.  The PID loops can accept input from either the control or monitoring thermometers, adding redundancy in the case of a thermometer failure.

In order to further buffer the radiation shield against radiative disturbances both within and exterior to the instrument, the radiation shield and \ln tank are wrapped in custom Multi-Layer Insulation (MLI) blankets. The blankets are made of alternating layers of highly reflective aluminized-Mylar sheets and a spacer material (nylon tulle) to prevent thermal shorting between the reflective mylar layers. The MLI blankets slow heat loss from the instruments' components, reducing demand on the TMC electronics. 

\begin{figure}
\begin{center}
\includegraphics[width=0.6\columnwidth]{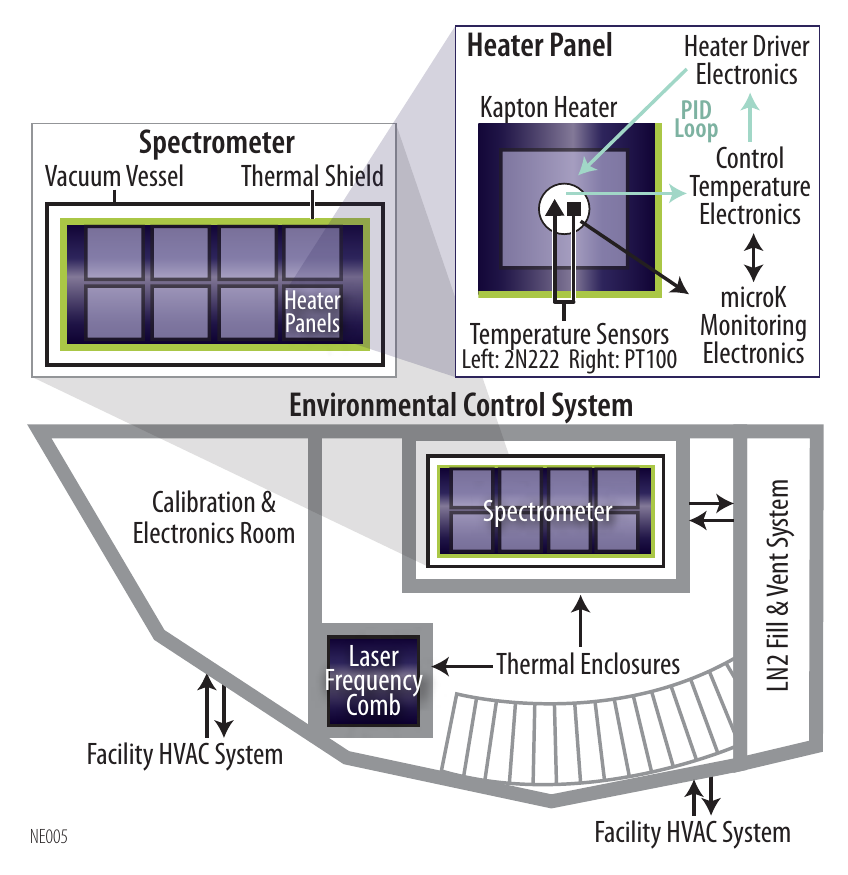}
\caption{\footnotesize Schematic of the hierarchical system of passive and active thermal control systems for NEID and its laser frequency comb as they will be assembled in the WIYN basement.  NEID and the LFC will be surrounded by passive food storage enclosures to dampen high-frequency temperature variability, while electronic heaters and thermometers actively control the volume surrounding the spectrometer optics.}
\label{fig:basement}
\end{center}
\end{figure}

The final element of the NEID/HPF ECS is a passive exterior thermal enclosure to minimize large, short-term temperature variations. Our passive enclosure is a commercial chamber from Bally Enclosures, commonly used for food storage. It consists of rigid inner and outer walls, with 4" insulating foam in between. NEID and its LFC are intended to be housed in separate enclosures to isolate the spectrometer from the heat loads created by the LFC and its associated electronics.  The configuration of the instrument, LFC, and their respective thermal enclosures in the WIYN basement is shown in Figure \ref{fig:basement}.  Note that the LFC enclosure may not be employed if the installed HVAC is sufficient to stabilize the environment which includes the LFC outside of the spectrometer enclosure.

\begin{figure}
    \centering
    \includegraphics[width=0.9\columnwidth]{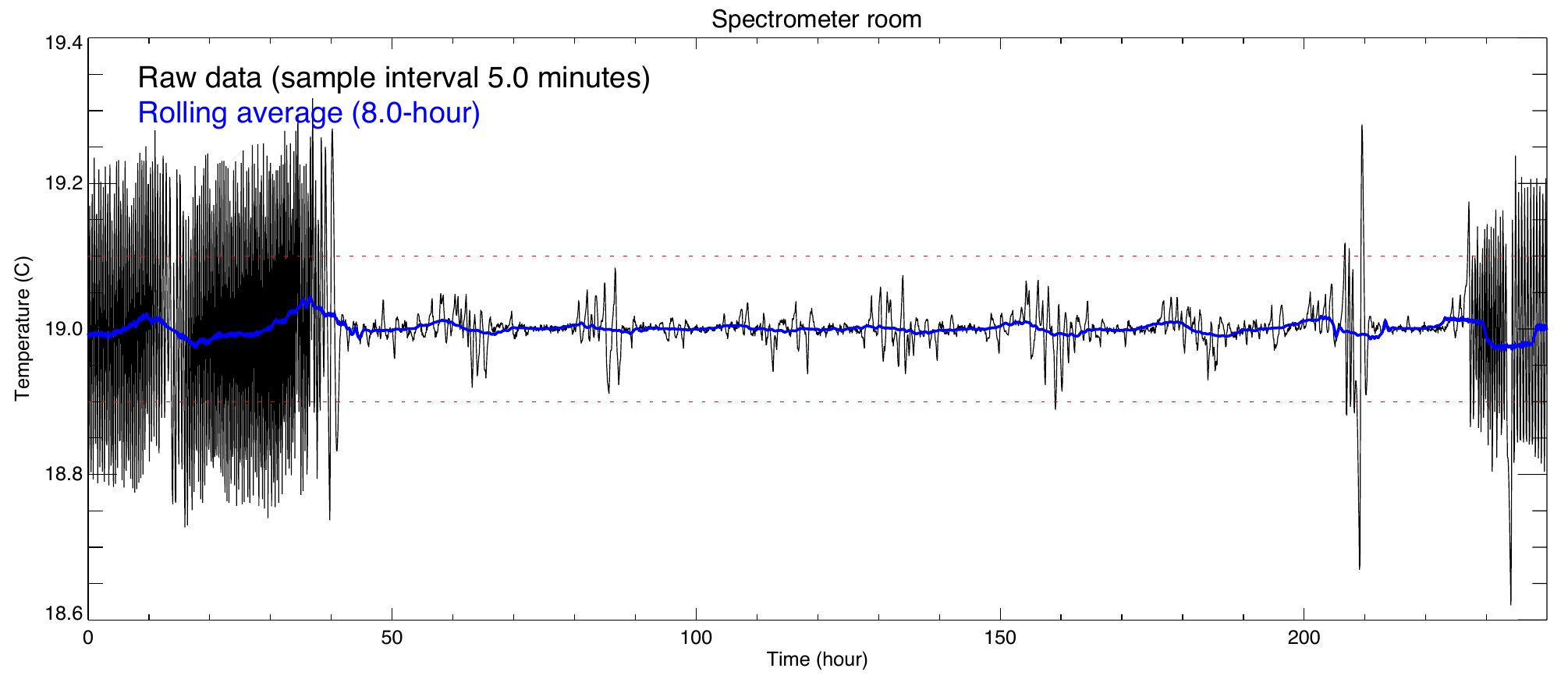}
    \caption{Results of a 10-day test of temperature stability inside the WIYN basement.  The change in short-term variability from hours $\sim$50-200 is believed to be caused by temperatures outside the WIYN basement dropping below $\sim5$~$^{\circ}$C, during which time the HVAC is more stable, but is still under investigation.  The system is compliant with our requirements throughout the demonstration.}
    \label{fig:basement_temps}
\end{figure}

Because our ECS is designed to stabilize the environment surrounding the optical bench, rather than actively controlling the temperature of the bench itself, the final equilibrium temperature of the bench is set by the black body cavity created by balancing the energy input from the TMC heaters and radiation output (ultimately) to the ambient environment outside the thermal enclosure.  Thus, the mean temperature of the ambient room must be kept near constant long-term ($\pm$0.2~$^{\circ}$C); changes to the mean temperature will lead to a gradual change in the bench's equilibrium temperature, regardless of the stability of the radiation shield.  Short duration temperature transients are filtered out by the thermal enclosure and TMC system. Setpoint maintenance may be achieved with HVAC systems commonly found in laboratory environments, but some care must be taken with, e.g., seasonal exchanges between heating and cooling modes.  In Figure \ref{fig:basement_temps}, we show the performance of the HVAC system in the WIYN basement over 10 days.  Our requirement for NEID demands that the 8-hour rolling average temperature must remain within $\pm 0.1$~$^{\circ}$C over the course of a month; Figure \ref{fig:basement_temps} shows performance meeting this requirement over a short timescale, and efforts to achieve similar stability long-term are ongoing.

\section{Differences Between NEID and HPF}
\label{sec:differences}
The core conceptual design for the ECS on both HPF and NEID is identical.  However, each system has instrument-specific requirements that prompted slight optimizations or alterations.  Here, we list the significant differences between the two instruments' environment control systems.

\begin{itemize}

\item \emph{Heat Sink}: The near-infrared bandpass of HPF required that its optics be cooled to 180 K.  Thus, the \ln tank on HPF served as that instrument's heat sink, in addition to cooling the detector and vacuum getters.  Copper straps connected the radiation shield to the \ln tank, and were sized to provide the cooling required for the optics to reach approximately 170 K. Because the optical bandpass of NEID is not hampered by thermal background radiation, it may be operated near room temperature, sparing us the complication and expense of installing thermal straps.  Instead, we use the TMC heaters to raise the temperature of the optics and radiation shield to 300 K, and use the entire ambient environment (293 K) as a uniform heat sink. In addition to minimizing thermal gradients across the active volume, this design decreases the \ln boil-off rate, reducing the magnitude of thermal transients caused by tank refills relative to HPF.

\item \emph{MLI Blanket Construction}: NEID operating temperature is above ambient and the NEID radiation shield is not thermally coupled to the \ln tank so the MLI blankets around the thermal shield play a different role in this instrument.  Thinner MLI blankets were used to \textit{increase} the ambient losses to approximately 1 watt per square meter which facilitated use as the primary heat sink; the NEID MLI blankets are 6 layers thick as opposed to 12 for HPF.
   
\item \emph{Radiation Shield Dimensions}:  The radiation shield's walls were made 50 percent thicker--or 3/16" thick--to minimize thermal gradients across the structure while remaining lightweight enough that a team of technicians can still handle its lid by hand.  These walls are made of the same high-conductivity 3003 aluminum as the HPF radiation shield to minimize thermal gradients.
 
\item \emph{Thermal Breaks}:  A 3/16" thick arch of 3003 aluminum, supported by brackets welded to the vacum chamber wall, was placed over the top of the \ln tank.  This passive shield further decoupled the \ln tank from the radiative environment viewed by the lower thermal shield to improve the stability of the thermal shield wall.

\item \emph{Monitoring Thermometers}: HPF operates at 180 K, which is outside the range of highest sensitivity (as defined by change in resistance per unit temperature change) for most resistive thermometers.  To achieve the highest temperature measurement precision possible, we equipped HPF with Cernox\texttrademark~1080 thermometers \cite{courts00} from Lake Shore Cryogenics, which typically exhibit higher sensitivity at 180 K than a standard platinum thermometer.  At the 300 K operating temperature of NEID, however, platinum resistive thermometers are usually more sensitive than Cernox\texttrademark~devices, and have the advantage of adhering to a standard calibration curve.  For NEID, we selected PT-103 thermometers from Lake Shore as our monitoring thermometers.

\item \emph{Heater Design and Placement}: As shown in S16, HPF's TMC heaters consist of a set of four chassis-mount resistive heaters mounted to an aluminum plate to distribute heat.  For NEID, we have instead used custom adhesive heaters from Birk Manufacturing. These heaters, shown in Figure \ref{fig:heater}, consist of a resistive wire woven through Kapton with vacuum-compatible (3M 966 PSA low outgassing) adhesive, which distribute heat more evenly across the active surface. Adhesive heaters were impractical for HPF, as Kapton with an adhesive layer becomes difficult to remove once cooled to cryogenic temperatures, but they are ideal for NEID. NEID uses 28 adhesive heaters across the surface of the radiation shield, twice as many as used for HPF.  The number and distribution of heaters reduces the amount of uncontrolled surface area across the radiation shield, and assigns a more uniform area to each controlled region.

Prior to building the NEID ECS, we simulated the performance of the modified ECS system using SolidWorks.  The thermal model was developed to help optimize the number, size, shape, and placement of the individual custom heaters and their corresponding control zones.  In the operating mode, each heater, temperature sensor, and corresponding PID loop acts against the varying exterior micro-environment that it ``sees" in its solid angle of radiative coupling.  More zones improve the precision of this response, but complicate the control and temperature sensing support needed.  NEID's design, including this model, helped to optimize this trade-off.  In Figure \ref{fig:gradient}a, we show the predicted temperature distribution across the radiation shield when the instrument is heated to its nominal (10K above ambient) operating temperature.  A uniform ambient environment was assumed for comparison purposes.  Our simulations predicted a thermal gradient of just 9 mK (peak-valley) across the radiation shield, and a gradient across the optical bench that was below the simulation's 0.1 mK precision. These values are more than 30 times less than predicted for the solid model of the HPF ECS.

\begin{figure}
\begin{center}
\includegraphics[width=0.5\columnwidth]{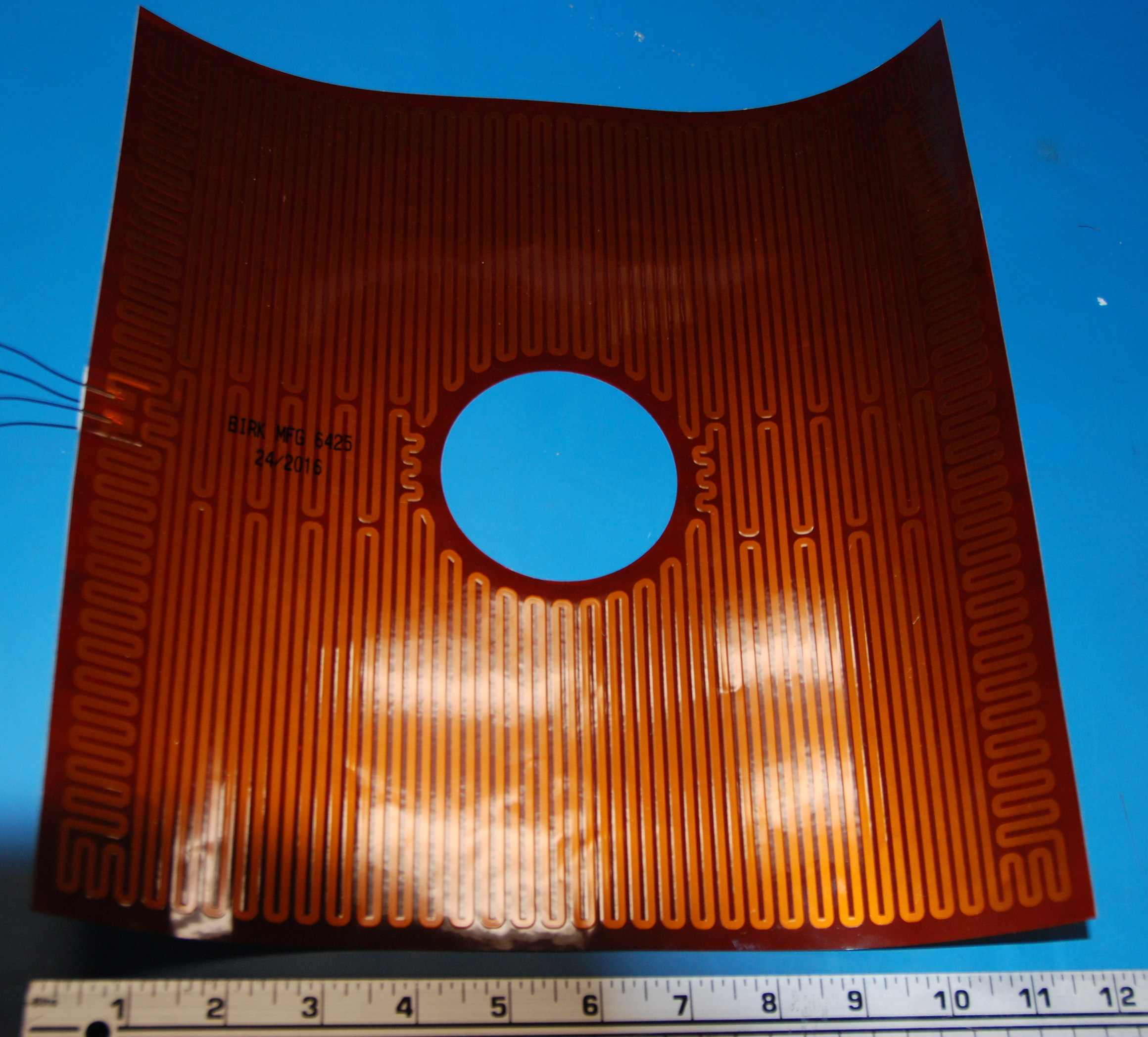}
\caption{\footnotesize A resistive heater used on the NEID radiation shield.  The heater is coated on one side with vacuum-compatible Kapton adhesive, which facilitates simple assembly and efficient heat distribution.}
\label{fig:heater}
\end{center}
\end{figure}

While the measurement \emph{precision} of our TMC system is exquisite, its absolute \emph{accuracy} is actually insufficient to measure the predicted temperature gradients for either HPF or NEID.  The accuracy of our thermometers is limited by the requirement that they be distributed throughout the radiation shield, resulting in long wire leads between the thermometers and the instrument's electronics cabinet.  Our 2N2222 thermometers use a 2-wire measurement system, wherein the current used to excite the device is sent through the same 2 wires used to measure the voltage across it.  Thus, the temperature-versus-voltage calibration curve changes as a function of wire length, and can only be accurately determined when the device is paired with a monitoring thermometer, as is the case for the control stations.  For 2N2222 thermometers installed farther from a control station, the absolute calibration may vary by as much as 0.5K.  Our PT-103 thermometers use a 4-wire system, which separates the wire leads used for delivering current and measuring voltage.  This allows for cancellation of voltage drops across the wires, and provides a more accurate measurement.  However, the actual device has only 2 leads, and the absolute calibration changes as a function of where the 4 wires are soldered to those leads.  Lake Shore's product specifications estimate a calibration offset of 9.5 mK for every millimeter away from the device that the wires are attached.  Thus, the expected variations in wire lead attachment will introduce calibration offsets larger than the small predicted gradients across the optical bench.

We emphasize, though, that the most important metric for precision Doppler velocimetry is not the absolute temperature gradient, but its variability.  Ultra-precise Doppler measurements are typically differential in nature, so while the absolute temperature gradient might influence the RV zero point, it is more important to eliminate velocity shifts induced by a varying thermal background or mechanical expansion/contraction caused by a variable gradient.

We are not yet able to validate the gradient variability for NEID, because we did not have enough thermometers distributed across the optical bench.  Space constraints within the radiation shield require that we place 2N2222 thermometers on the optics mounts rather than at many locations on the actual table, and NEID optics had not been installed as of this stability demonstration.  However, we can demonstrate the stability of the gradient for HPF, thereby confirming the performance of the overall conceptual design.  In Figure \ref{fig:gradient}b, we show the temperature difference between the two 2N2222 thermometers most distant from each other on the HPF bench.  The gradient has an RMS stability less than 0.5 mK, and is periodically driven by the daily filling of the \ln tank.  Because of the improved performance of the NEID ECS, and because of the greatly reduced thermal contact between NEID's bench and its \ln tank, we expect its temperature gradient to be much more stable.

\end{itemize}

\begin{figure}
\begin{center}
\includegraphics[width=\columnwidth]{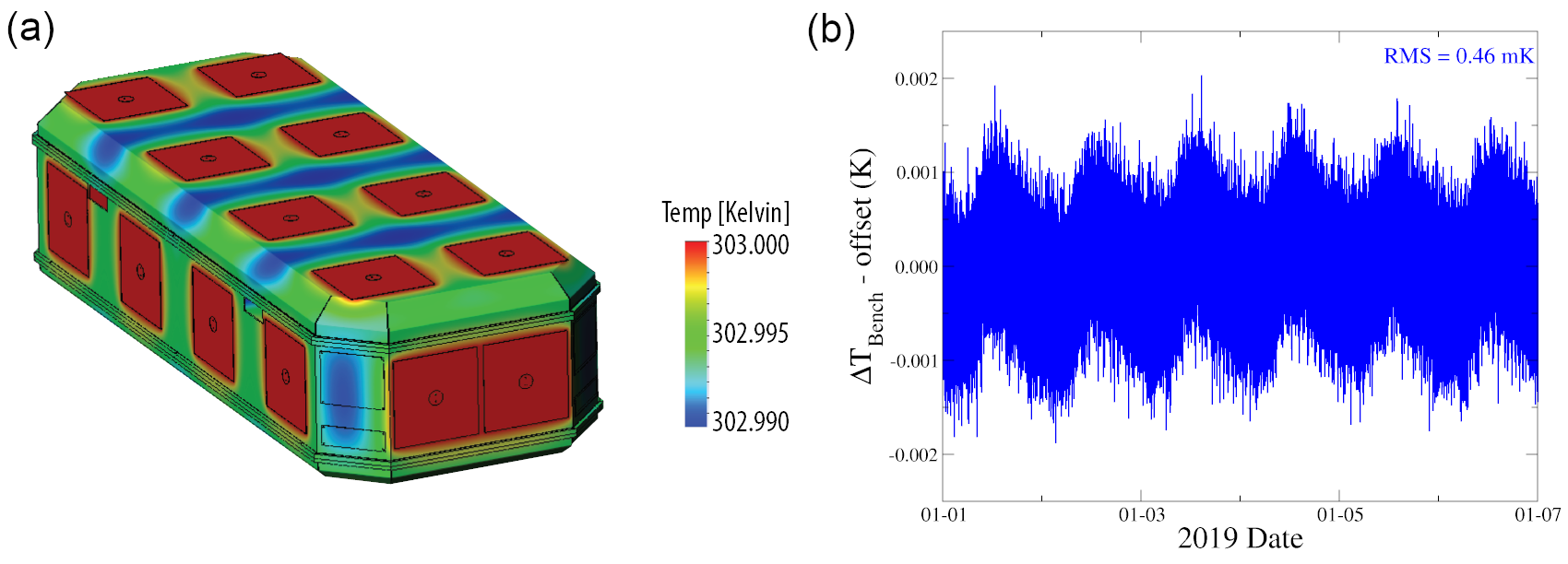}
\caption{\footnotesize \emph{a}: Predicted temperature distribution of the NEID radiation shield when the instrument is heated to its nominal 303K operating temperature.  The predicted gradient is just 9 mK peak-valley.  \emph{b}: Temperature difference between two optics across from each other on the HPF bench.  The RMS variability between these locations is better than 0.5 mK, confirming the stability of the temperature gradient for HPF.}
\label{fig:gradient}
\end{center}
\end{figure}

\section{TMC Electronics}
\label{sec:electronics}

A significant change we have implemented on both HPF and NEID since the publication of S16 is the final version of our custom electronics to monitor and control the temperature of the radiation shield.  Our TMC electronics use the same basic circuit design and operational scheme as those used for the stability demonstration from S16.  They have been upgraded to improve their durability, to eliminate sources of drift identified in early tests, and to shield against sources of electrical noise commonly found in observatory settings.  Here, we describe our TMC electronics in detail for the first time.

\begin{figure}
\begin{center}
\includegraphics[width=\columnwidth]{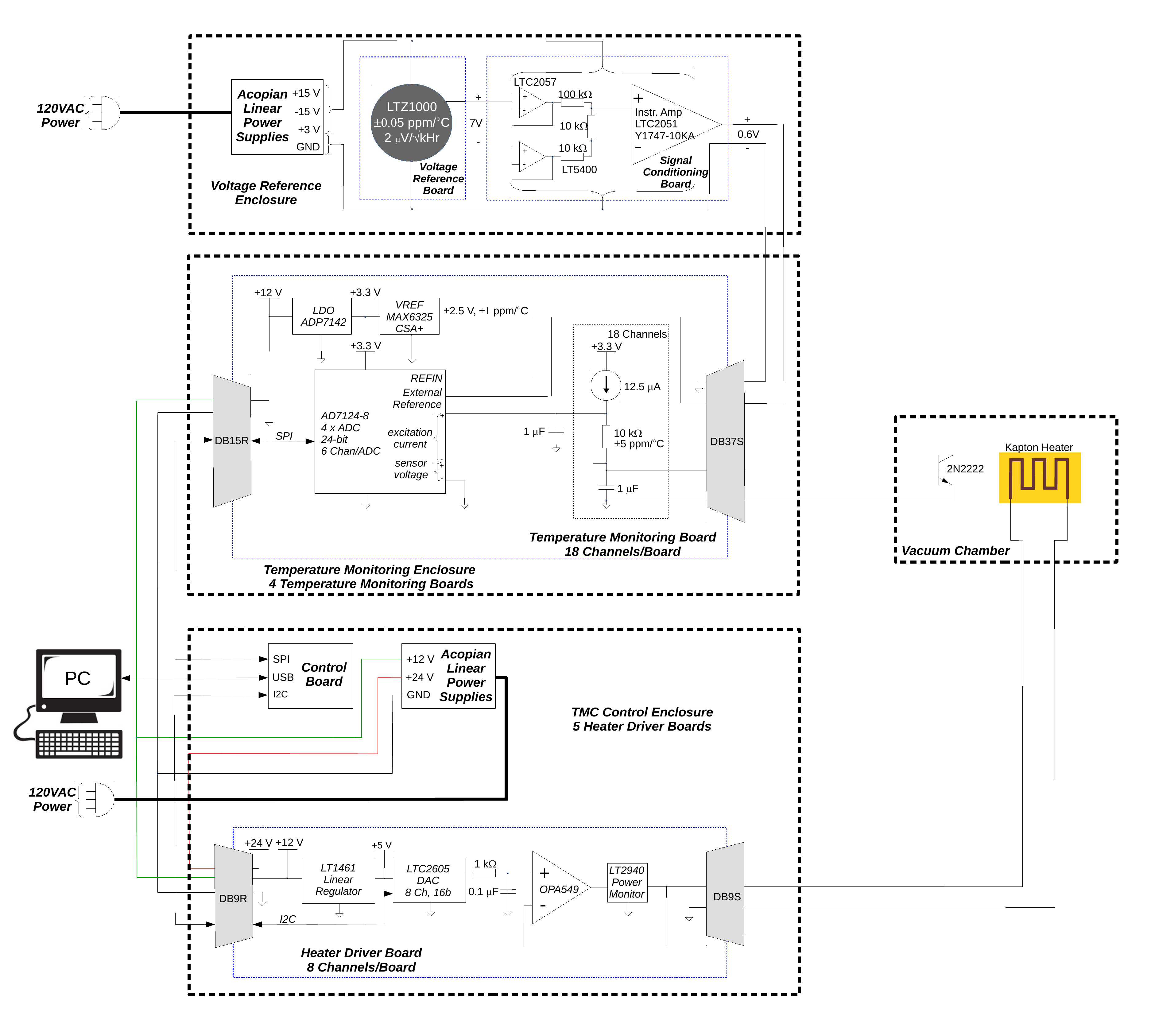}
\caption{\footnotesize Simplified schematic diagram of the TMC electronics.}
\label{fig:tmc_electronics}
\end{center}
\end{figure}

Figure \ref{fig:tmc_electronics} shows a simplified schematic for the TMC electronics. TMC circuitry is divided between three separate 19-inch, rack-mounted enclosures: the Temperature Monitoring Enclosure, the Voltage Reference Enclosure, and the Temperature Control Enclosure.

The Temperature Monitoring Enclosure contains four Temperature Monitoring Boards, each with the capability of reading out 18 channels of diode-connected 2N2222 temperature sensors. Each channel contains a temperature-compensated 12.5~$\mu$A current source that provides excitation current to the 2N2222 transistors. Each board contains three AD7124-8, 24-bit, sigma-delta ADCs that measure the sensor voltage, which is nominally 0.5~V at room temperature with a temperature coefficient of approximately -2.2~mV/$^{\circ}$C. Use of such a high dynamic range ADC allows the TMC to utilize a single gain setting over the full 77~K to 320~K temperature range spanned by HPF and NEID, while still maintaining a measurement resolution of $<100$~$\mu$K. The ADCs also monitor sensor excitation current by measuring the voltage across a 10~k$\Omega$, 5~ppm/$^{\circ}$C precision resistor.  The Temperature Monitoring Boards offer rapid temperature monitoring; we can measure the voltages of up to 72 2N2222 thermometers on a 6-second cadence.  For comparison, our microK 250 reads a set of 30 PT-103 thermometers on a 50-second cadence.

TMC stability is maintained via comparison with an ultra-low drift, 0.6 V reference voltage created within the Voltage Reference Enclosure. Internally, the Voltage Reference Board utilizes a $\sim$7~V, 0.05~ppm/$^{\circ}$C ultra-stable reference level from the ovenized, LTZ1000 buried-zener reference. This reference level is sent to the Signal Conditioning Board, wherein it is buffered by the LTC2057 zero-drift amplifier and divided down to $\sim$0.6~V through the 10~k$\Omega$/100~k$\Omega$ matched resistor network of the LT5400-3. This final $\sim$0.6~V reference level is provided to the Temperature Monitoring Boards through an instrumentation amplifier formed from the LTC2051 zero-drift amplifier and the 10 k$\Omega$/10 k$\Omega$ Y1747-10KA matched resistor network. All components were selected for ultra-low temperature drift and excellent long-term stability. The final reference voltage demonstrated $<0.2$~$\mu$V/$^{\circ}$C temperature drift when externally cycled at a rate of $\sim\pm$1~$^{\circ}$C/min over a 5$^{\circ}$C to 30$^{\circ}$C range. In the final assembly, the Voltage Reference Board and Signal Conditioning Board were surrounded by weather stripping material in order to insulate them from local air currents and short-term changes in ambient temperature.

The Temperature Control Enclosure includes five Heater Driver Boards, each of which provides 8 channels of up to $\sim$3~W of drive for each Kapton heater; normal operation requires only a small fraction of this power. The heater driver design is identical to that used for the HPF system presented in S16, and utilizes the OPA549 high-power amplifier to servo on a programmable power level. Power output is monitored via the LT2940 four-quadrant power monitor. The Heater Driver Boards mount in a card cage inside of the Temperature Control Enclosure. Air-flow within the enclosure is designed so as to force air between the boards, with fans mounted in the rear of the unit. 

The Temperature Control Enclosure also includes a small NIOS2-based digital control board, which manages the SPI readout of the Temperature Monitoring Boards, the I2C control of the Heater Driver Boards, and the USB connection to the main Science Computer. The Heater Driver Boards and Temperature Monitoring Boards are powered via Acopian linear power supplies (+24~V, +12~V) mounted inside of the Temperature Control Enclosure. The Voltage Reference Board and Signal Conditioning Board are powered via Acopian linear power supplies ($\pm$15~V, +3~V) within the Voltage Reference Enclosure. All design documents, including schematics, bill-of-materials, and PCB layout files, are archived as a resource to the community\footnote{https://scholarsphere.psu.edu/collections/3xs55m982m}.  

\section{Description of Stability Test}
\label{sec:test}

In S16, we performed an environment stability test using the HPF cryostat operated in a ``warm'' configuration as a proof of concept for the NEID system.  That test demonstrated more than sufficient stability for the design specifications of HPF, but does not reflect the performance expected for NEID after the implementation of the NEID-specific adaptations and the upgraded TMC electronics.  Thus, we have conducted a full-scale stability test using the actual NEID vacuum chamber to verify its performance in a laboratory setting.

Our NEID stability test was conducted between March and May of 2018 in our spectrometer integration laboratory at Penn State University's Innovation Park campus. The vacuum chamber was evacuated and heated with the optical bench in place, but without the optics and mounts.  To simulate conditions as similar as possible to the WIYN basement, we sealed the vacuum chamber in its passive thermal enclosure during this test.

The stability of the NEID environment was tracked by two primary observables.  To evaluate vacuum stability, we monitored the pressure inside the vacuum chamber using the pressure reading from two dedicated MKS MicroPirani vacuum gauges, along with the pressure reading from the NexTorr D-100 ion pump. To monitor temperature stability, we mounted two identical PT-103 thermometers separated by approximately 1 inch in the center of the optical bench. These thermometers were read out by the microK 250 temperature monitor.

Because our objective is to dampen or prevent changes in ambient temperature propagating through to the optical bench, it is instructive to carefully monitor the ambient conditions during an ECS performance test.  To that end, we placed several adhesive-mounted PT-100 thermometers from Omega Engineering around our laboratory and inside the thermal enclosure.  These thermometers were monitored by a Model 218 temperature monitor from Lakeshore Cryotronics.

\section{Results of Stability Test}
\label{sec:data}

\subsection{Vacuum Stability}
\begin{figure}
\begin{center}
\includegraphics[width=0.8\columnwidth]{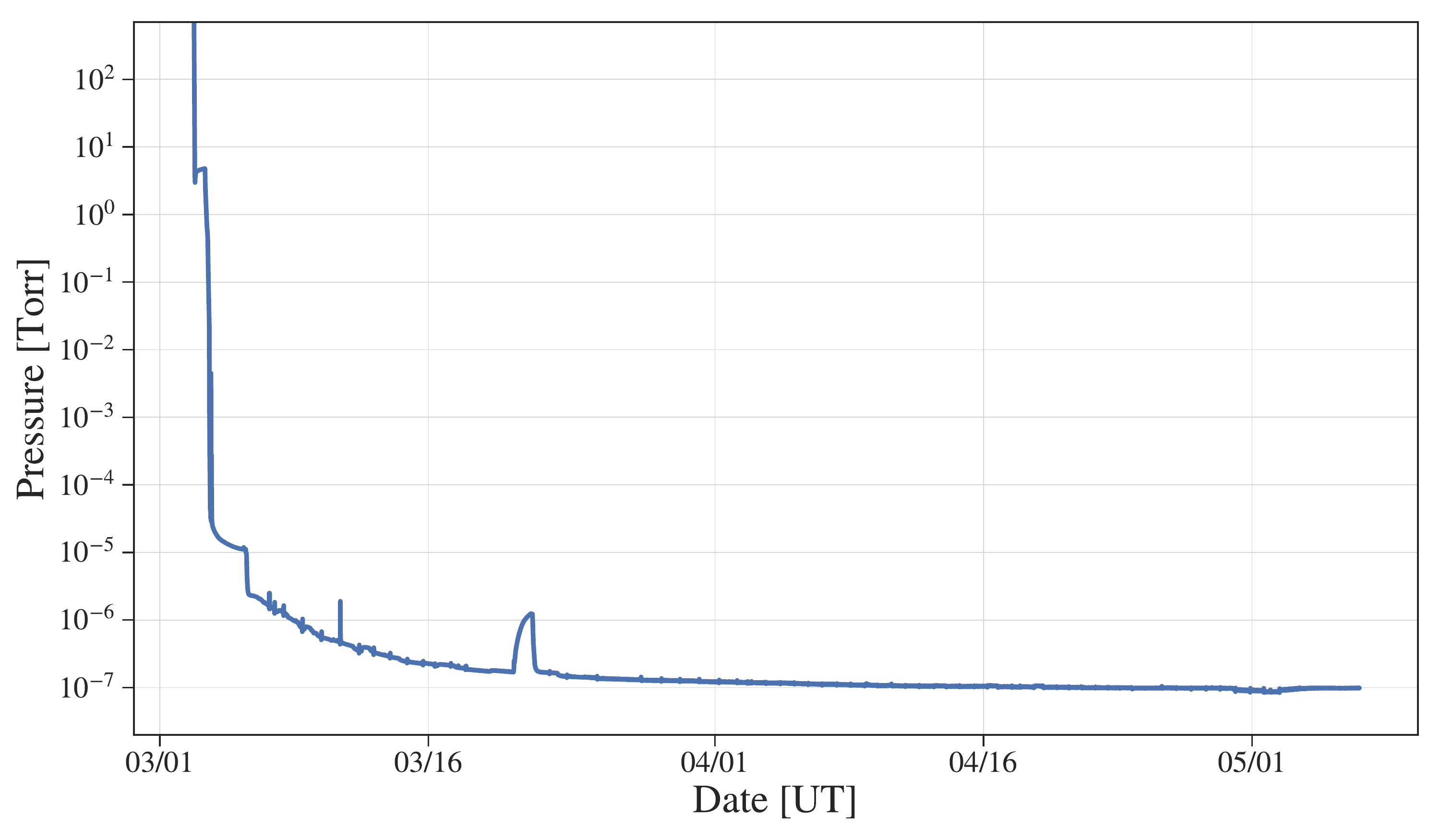}
\caption{\footnotesize NEID's long term pressure stability. Pressure is brought from ambient pressures at the start of the stability run in early March 2018, and stabilized at the $\sim10^{-7} \unit{Torr}$ level. The short pressure spike on March 22nd was due to a controlled experiment to study the hydrogen outgassing rate in the NEID spectrometer (see further description in text).}
\label{fig:pressure}
\end{center}
\end{figure}

Figure \ref{fig:pressure} shows the pressure as recorded by one of the NEID vacuum gauges during the stability test. We only show the vacuum reading from one of the 3 NEID vacuum gauges for clarity, as the other two vacuum gauges all demonstrated a consistent pressure level within a factor of 2-3. At the beginning of the stability run in early March, the pressure was brought down from ambient pressures levels to the $\sim 5 \unit{Torr}$ level using a Tri-Scroll 300 I Phase vacuum pump, and down to the $\sim10^{-5} \unit{Torr}$ level using an Agilent TwisTorr 84 Turbo-molecular pump (using the Tri-Scroll 300 pump as a backing pump). Ultimate pressures were reached by filling the LN2 tank to activate the charcoal getters mounted on the sides of the NEID LN2 tank along with activating the NexTorr hydrogen getter and ion pump. After about 1 month since the start of the stability run (early April) the vacuum chamber maintained a $1 \times 10^{-7} \unit{Torr}$ pressure level or better for the rest of the stability run of over 1 month. Comparing directly to S16, they demonstrated slightly higher quality vacuum achieved with HPF than this---or vacuum pressures below $10^{-7}$ Torr over 2 months. This is expected, given that NEID is mostly a warm instrument operated at 300K. Even so, as discussed in S16, a vacuum stability of $<1e-6 \unit{Torr}$ is sufficient to minimize thermal conduction in the air within the spectrometer, as well as reducing the RV errors due index of refraction variations due to pressure fluctuations within the instrument to the $\sim0.02\unit{cm/s}$ level.

The short pressure spike seen in Figure \ref{fig:pressure} on March 22 was due to a controlled 24 hour experiment where the NexTorr hydrogen getter and ion pump were valved out from the NEID vacuum chamber. The goal of this experiment was to study the hydrogen outgassing rate from the NEID vacuum chamber walls, but hydrogren is soluble in stainless steel where it can get trapped in precipitates dueing fabrication, slowly outgassing at room temperature \cite{vacuumhandbook}. In 24 hours the vacuum pressure rose from $1.7 \times 10^{-7} \unit{Torr}$ to $1.2 \times 10^{-6} \unit{Torr}$, corresponding to an outgassing rate of $1.03 \times 10^{-6} \unit{Torr/day}$. This outgassing rate is similar to the outgassing rate seen with HPF which slowly decayed from $\sim1.0 \times 10^{-6} \unit{Torr/day}$ to the $1.0 \times 10^{-7} \unit{Torr/day}$ level over 9 months (see Figure 10 in S16). After valving in the NexTorr hydrogen getter, the vacuum pressures quickly returned to the previous baseline vacuum level and stabilized at the $10^{-7}$ Torr level after $\sim$1 month.

\subsection{Temperature Stability}

\begin{figure}
\begin{center}
\includegraphics[width=0.7\columnwidth]{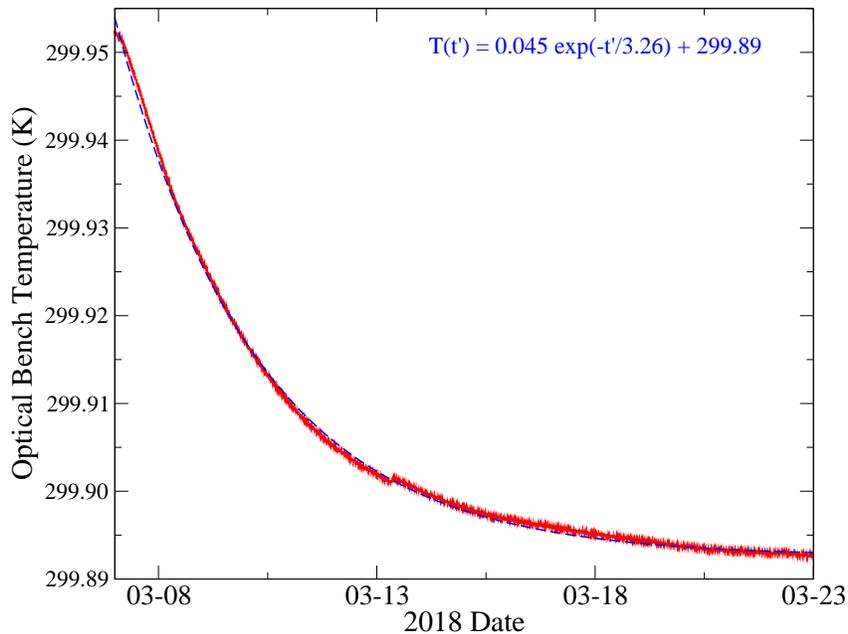}
\caption{\footnotesize Temperature of the NEID optical bench after an initial warming using the warm-up heaters.  The temperature (\emph{red}) decayed asymptotically towards equilibrium.  An asymptotic exponential model (\emph{blue}) yields a decay time constant of approximately 3 days.}
\label{fig:decay}
\end{center}
\end{figure}

One logistical consequence of our high-quality vacuum is that in addition to facilitating long-term temperature stability, it requires a long time for the optical bench to settle to equilibrium.  While we use an array of dedicated warm-up heaters (offering nominally 125W input power) mounted to the bottom of the NEID optical bench to drive the bench near its equilibrium value, it is impossible to ``set'' it at the correct temperature to sub-milliKelvin precision. Thus, at the start of any vacuum cycle we must wait for the bench temperature to asymptotically approach equilibrium.  In Figure \ref{fig:decay}, we show the bench temperature, measured using the microK/PT-103 system, as it settled to equilibrium during our stability test.  The small temperature increase on March 13 is the result of the PID loops on the radiation shield heaters resetting during a switch between PT-103 and 2N2222 thermometers as the active control sensors.  We find that the temperature during this period is well described by an asymptotic exponential model, $T(t) = Ce^{-t/\tau} + T_0$.  Our system typically yields a time constant $\tau \sim 3$ days, which means that if we define temperature ``stability'' as the middle bench temperature sensor showing a derivative of $|\frac{\delta T}{\delta t}| < 1$ mK/week, the bench must be allowed to settle for approximately 1 month.

\begin{figure}
\begin{center}
\includegraphics[width=0.8\columnwidth]{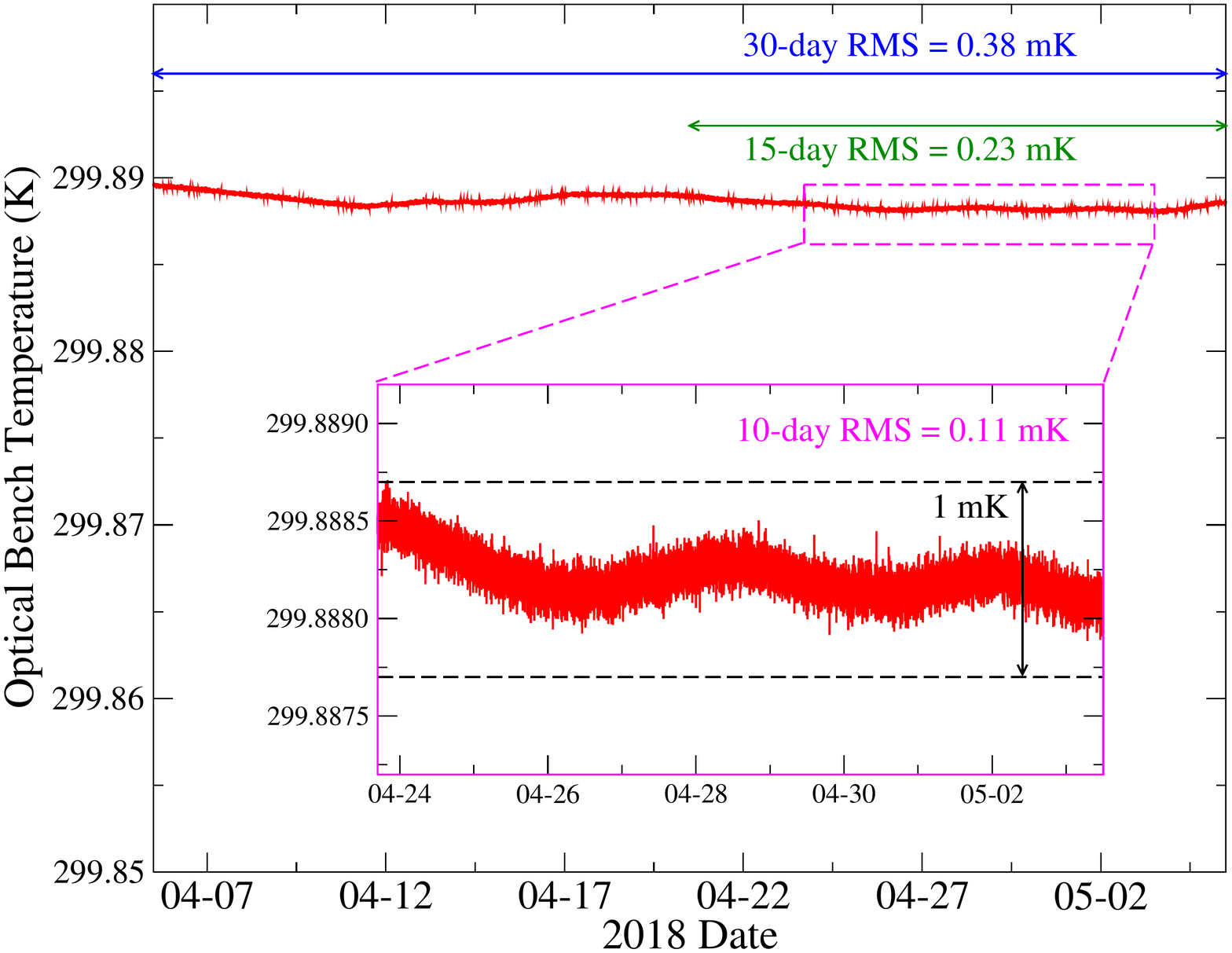}
\caption{\footnotesize Temperature of the NEID optical bench during the 30-day stability trial.  An inset is shown for a representative 10-day sample, with low-level variability caused by the daily \ln fill.}
\label{fig:stability}
\end{center}
\end{figure}

Once the bench temperature reached equilibrium, we left the system undisturbed except for daily filling of the \ln tank.  The resulting temperature stability is shown in Figure \ref{fig:stability}.  For visual clarity, we only show the temperature as recorded by one of the two PT-103 thermometers mounted on the bench; the two thermometers showed identical stability and noise structures, but with a zero-point offset consistent with the calibration uncertainties described in \S \ref{sec:differences}.  For the entire 30-day experiment, we observed an RMS temperature stability of 0.38 mK, with any 10-day period having an RMS scatter of 0.10-0.15 mK.

\begin{figure}
\begin{center}
\includegraphics[width=\columnwidth]{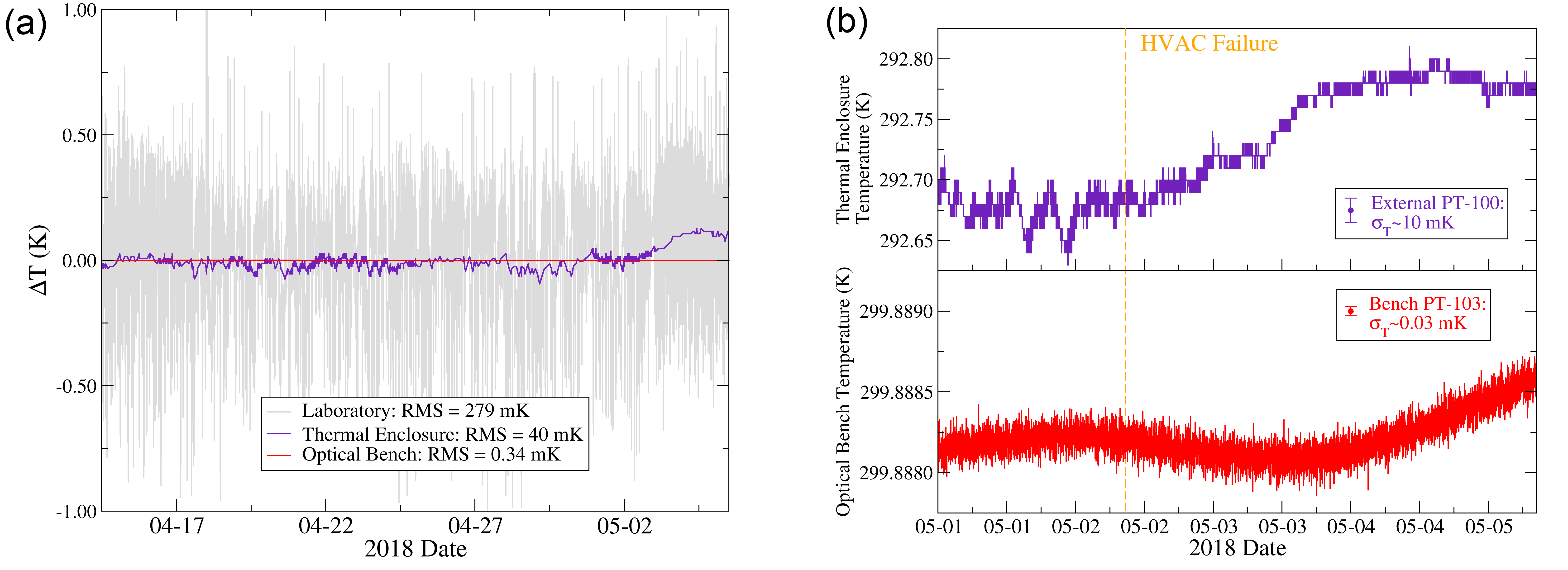}
\caption{\footnotesize \emph{a}: Temperature deviations from the mean for our laboratory (\emph{gray}), the inside of the thermal enclosure (\emph{purple}), and the optical bench (\emph{red}).  \emph{b}: Inset of the last days of our test, when an HVAC failure (\emph{orange line}) caused the ambient temperature to increase.  The (time-delayed) warming of the optical bench shows that the bench temperature is coupled to the mean external temperature.  Representative single-measurement uncertainties are shown for the external PT-100 and internal PT-103 thermometers, but note the different Y-axis scales.}
\label{fig:layers}
\end{center}
\end{figure}

The small variability present in Figure \ref{fig:stability} is dominated by temperature changes in the laboratory.  In Figure \ref{fig:layers}, we show the relative temperature stability of our instrument development lab, thermal enclosure, and optical bench.  While the bench temperature is still measured using the microK device, the thermal enclosure and instrument lab are monitored using adhesive-mounted PT-100 thermometers from Omega Engineering, and read out using Model 218 temperature monitor from Lake Shore Cryogenics.  The Model 218 was read out every 10 seconds, as opposed to the 50-second cadence of the microK 250.  While the ECS is immune to high-frequency variability such as produced by commercial HVAC systems, the bench temperature will drift in response to changes in the mean laboratory temperature (reflected by the thermal enclosure temperature) over days or weeks.  Maintaining a constant mean temperature in the WIYN basement is a facility requirement for NEID, but the HVAC system at our spectrometer development lab is less robust. Hence, we expect the NEID ECS to be even more stable when deployed at WIYN (Figure \ref{fig:basement_temps}).  Based on this experiment, we can draw conclusions for the relative performance of the passive and active components of our ECS.  Using just the passive thermal enclosure (RMS = 40 mK), we achieve approximately an order of magnitude in improvement over the laboratory HVAC (RMS = 279 mK).  Within the actively-controlled volume, the temperature is another 2 orders of magnitude more stable.  Overall, this stability test shows a factor of $\sim700$ reduction in thermal variability from the ambient environment to the optical bench, exceeding the factor of 200 assumed in our error budget \cite{halverson16}.

\section{Discussion}
\label{sec:discussion}

The environmental stability of NEID is, as expected, a significant improvement over that of HPF.  Comparing directly to S16, their demonstration on HPF yielded vacuum pressures below $10^{-7}$ Torr over 2 months, while NEID is expected to operate a slightly lower quality vacuums (nominally better than $10^{-6}$ Torr) since it is mostly a warm instrument.  With regards to temperature stability, the NEID system achieved $<0.4$ mK stability over 30 days, whereas HPF exhibited $0.6$ mK over 15 days in both warm and cold configurations.  As shown in Figure \ref{fig:stability}, the best 15-day period for NEID had an RMS temperature scatter of just $0.23$ mK, nearly a factor of 3 better than the 15-day HPF trial.

It is important to emphasize that neither this experiment nor that of S16 represents the ultimate performance of either instrument.  We have deployed the revised TMC electronics described herein on both instruments, which has resulted in a significant improvement in HPF's stability.  Furthermore, we are optimistic that upgrades to the WIYN basement will provide more stable mean temperatures than achievable in our laboratory, which should eliminate some of the low-frequency drifts observed here.

As emphasized by S16, our ECS is highly flexible and modular, making it applicable to instruments of various scales, budgets, and applications.  The complete vacuum chamber and ECS described here is relatively inexpensive, with a total cost of approximately \$500\,000 USD, and could be adapted to an even more modest budget.  Significant cost savings could be achieved by adopting a smaller commercial vacuum chamber \cite{hearty14}, and omitting some of the components we use to achieve ultra-stable conditions such as the redundant temperature monitoring system and the ion hydrogen getter.  To that end, we have made the design of our ECS completely open-source, beginning with the cryostat design released in S16, and now including the design of our TMC electronics. 

The performance of the NEID ECS is a testament to the value of heritage in design of astronomical instrumentation.  The design of our vacuum chamber is adapted from that of APOGEE \cite{blank10}, and its adaptation to the actively-controlled system originally developed for HPF was a years-long process.  By essentially copying the HPF ECS with minimal modifications, we were able to build and test a fully-operational system---whose performance already exceeds that of its predecessors---within about 6 months.  The ability to develop the critical components of NEID in such a short time frame is essential for the instrument to be ready to characterize new exoplanets discovered by TESS.

\section{Summary}
We have presented a full-system demonstration of the environment control system for the ultra-precise NEID Doppler spectrometer. During a 30-day stability test the NEID ECS achieved a temperature stability better than 0.4 mK RMS and a maintained a vacuum level better than $10^{-6} \unit{Torr}$, illuminating the path to achieving 10cm/s RV precision. Our system is modular and open-source, and can be scaled and applied to any number of instruments and purposes.

\acknowledgments 
NEID is funded by JPL under contract 1547612.  This work was partially supported by funding from the Center for Exoplanets and Habitable Worlds. The Center for Exoplanets and Habitable Worlds is supported by the Pennsylvania State University, the Eberly College of Science, and the Pennsylvania Space Grant Consortium.  PR and SH contributed to this work in part under contract with the California Institute of Technology(Caltech)/Jet Propulsion Laboratory (JPL) funded by NASA through the Sagan Fellowship Program executed by the NASA Exoplanet
Science Institute. GKS wishes to acknowledge support from NASA Headquarters under the NASA Earth and Space Science Fellowship Program-Grant NNX16AO28H. We acknowledge support from NSF grants AST-1006676, AST-1126413, AST-1310885, the NASA Astrobiology Institute (NAI; NNA09DA76A), and the Penn State Astrobiology Research Center.

\emph{Software:} \texttt{Jupyter} \cite{jupyter2016}, \texttt{matplotlib} \cite{hunter2007}, \texttt{numpy} \cite{vanderwalt2011}, \texttt{pandas} \cite{pandas2010}, SolidWorks, Zemax OpticStudio.


\bibliography{report}   
\bibliographystyle{spiejour}   


\end{document}